\documentstyle[12pt,subeqnarray]{article}

\begin{document}
\newcommand{\dis}{\displaystyle}
\newcommand{\dml}{{\em dim ~ ~ ker ~ ~}}
\newcommand{\expon}{{\rm e}}
\newcommand{\id}{ 1 \hspace{-2.85pt} {\rm I} \hspace{2.5mm}}
{\thispagestyle{empty}
\rightline{UT-702,1995} 
\rightline{NUS-HEP-95-03}
\rightline{} 
\rightline{} 

\centerline{\large \bf Q-Deformed Oscillator Algebra}
\centerline{\large \bf and an Index Theorem for the Photon Phase Operator}


\vskip 0.2cm
\centerline{ Kazuo Fujikawa {\footnote{E-mail
address: fujikawa@danjuro.phys.s.u-tokyo.ac.jp}}}
\centerline{ \it Department of Physics, University of Tokyo,}
\centerline{ \it Bunkyo-ku, Tokyo 113, Japan }
\centerline{L. C. Kwek {\footnote{E-mail address: scip3057@nus.sg}},
and C. H. Oh {\footnote{E-mail address:
phyohch@nus.sg }}
}
\centerline{{\it Department of Physics, Faculty of Science, } }
\centerline{{\it National University of Singapore, Lower Kent Ridge,} }
\centerline{{\it Singapore 0511, Republic of Singapore. } }}
\vskip 0.1in


\vskip 0cm

\centerline{\bf Abstract}

\noindent{The quantum deformation of the oscillator algebra and its
implications on the phase operator are studied from
 a view point of an index theorem by using an explicit matrix
representation.  For a positive deformation parameter $q$ or
$q=exp(2\pi i\theta)$ with an irrational $\theta$, one obtains an index
condition $\dml a - \dml a^{\dagger} = 1$ which allows only a
non-hermitian phase operator with $\dml \expon^{i \varphi} - \dml
(\expon^{i\varphi})^{\dagger} = 1$.  For $q=exp(2\pi i\theta)$ with a
rational $\theta$ , one formally obtains the singular situation $\dml a
=\infty$ and $ \dml a^{\dagger} = \infty$, which allows a hermitian
phase operator with $\dml \expon^{i \Phi} - \dml
(\expon^{i\Phi})^{\dagger} = 0$  as well as the non-hermitian one with
$\dml \expon^{i \varphi} - \dml (\expon^{i\varphi})^{\dagger} = 1$.
Implications of this interpretation of the quantum deformation are
discussed. We also show how to overcome the problem of negative norm
for $q=exp(2\pi i\theta)$. }\

(To be published in Modern Physics Letters A)

\newpage

\section{Introduction}
The presence or absence of a hermitian phase operator for the photon is
an old and interesting problem \cite{dirac,suss,pegg}, see ref
\cite{carr} for earlier works on the subject.  Recently, one of the
present authors \cite{fuji} introduced the notion of index into the
analysis of the phase operator.  The basic observation is that the
creation and annihilation operators of the oscillator algebra

\begin{equation}
[ a, a^{\dagger} ] = 1
\end{equation}

\noindent satisfy the index condition:
\hspace{-2mm}\renewcommand{\thefootnote}{\fnsymbol{footnote}}{\footnote[4]{The
index of a linear operator $a$, for example, is defined as the number
of normalizable states $u_{n}$ which satisfy $a u_{n} = 0$.}}

\begin{equation}
\dml a - \dml a^{\dagger} = 1 \label{rel2}
\end{equation}

\noindent as seen from the conventional representation

\begin{equation}
a = |0 ><1| +  |1 ><2|\sqrt{2} + |2 ><3|\sqrt{3} + \cdots .
\end{equation}

The state vectors $|k >$ are defined by $N |k > = k |k>$, where $N$ is the
number operator.  The phase operator defined by \cite{suss}

\begin{eqnarray}
\expon^{i \varphi} & = & \frac{1}{\sqrt{N+1}}a\nonumber\\
                   & = & |0 ><1| +  |1 ><2| + |2 ><3| + \cdots \label{rel4}
\end{eqnarray}

\noindent faithfully reflects the index relation (\ref{rel2})

\begin{equation}
\dml \expon^{i \varphi} - \dml ({\expon^{i \varphi}})^{\dagger} = 1.
\label{rel5}
\end{equation}

\noindent On the other hand, if one assumes a polar decomposition $a =
U( \phi) H$ with a unitary $U( \phi)$ and a hermitian $H$, one
inevitably has

\begin{equation}
\dml a - \dml a^{\dagger} = 0 \label{rel6}
\end{equation}

\noindent since the action of the unitary operator $U( \phi)$ is simply
to re-label the names of the basis vectors.  From these considerations,
one concludes that the phase operator $\varphi$ in (\ref{rel4})
cannot be hermitian, i.e., $e^{i\varphi}$ is not unitary.  A truncation of the
representation space of $a$
to $(s + 1) \times (s + 1)$ dimensions, however, generally leads to the
index relation (\ref{rel6})[5], and thus an associated phase operator
$\phi$ could be hermitian.  In fact, a hermitian phase operator $\phi$
may be defined by \cite{pegg}

\begin{equation}
\expon^{i \phi} = |0 ><1| +  |1 ><2| + |2 ><3| + \cdots + |s - 1><s| +
\expon^{i \phi_{0}}|s><0|   \label{rel7}
\end{equation}

\noindent with a positive integer $s$ ( a cut-off parameter) and an
arbitrary constant $\phi_{0}$.  The unitary operator $\expon^{i \phi}$
naturally satisfies the index condition

\begin{equation}
\dml \expon^{i \phi} - \dml ({\expon^{i \phi}})^{\dagger} = 0, \label{pegg1}
\end{equation}

\noindent and gives rise to a truncated operator

\begin{eqnarray}
a_{s} &=& \expon^{i \phi} \sqrt{N}  \nonumber \\
&= & |0 ><1| +  |1 ><2|\sqrt{2} + |2 ><3|\sqrt{3} + \cdots  |s - 1><s|
\sqrt{s} \label{pb2}
\end{eqnarray}

\noindent with

\begin{equation}
\dml a_{s} - \dml a_{s}^{\dagger} = 0
\end{equation}

\noindent since ~ $ ~ a_{s}^{\dagger}|s> = 0$.

The index relations (\ref{rel5}) and (\ref{pegg1}) clearly show the
unitary inequivalence of $\expon^{i \varphi}$ and $\expon^{i \phi}$
even for arbitrarily large $s$. Since the kernel of $a_{s}^{\dagger}$
is given by $ker\ a_{s}^{\dagger} = \{ |s\rangle \}$ in (10), which is
ill-defined in the limit $s \rightarrow \infty$, we analyze the behavior of
$e^{i\phi}$ for sufficiently large but finite $s$. To make this statement of
 large $s$ meaningful, we need to introduce a typical number to characterize
a physical system, relative to which the number $s$ may be chosen much
larger.
We thus  expand a physical state as

\begin{equation}
|p > = \sum_{n = 0 }^{\infty} p_{n} |n >.
\end{equation}

\noindent The finiteness of $<p | N^{2} |p >$ requires

\begin{equation}
 \sum_{n } n^{2} ~|p_{n}|^{2} = N_{p}^{2} < \infty \label{rel12}
\end{equation}

\noindent in addition to the usual condition of a vector in a Hilbert space,

\begin{equation}
\sum_{n } | p_{n} |^{2} < \infty .
\end{equation}

\noindent The number $N_{p}$ in (12) specifies a typical number associated to
a given physical system $|p\rangle$. By choosing the parameter $s$ at
$s >> N_{p}$, one may analyze the physical implications of the state
$|s\rangle$ , which is responsible for the index in (10), on the physically
observable processes.
It was shown in \cite{fuji} that the origin
of the index mismatch between (\ref{rel4}) and (\ref{rel7}), namely
the state $|s\rangle$ in (7),   is also
responsible for the absence of minimum uncertainty states for the
hermitian operator $\phi$ in the characteristically quantum domain with
small average photon numbers.

A major advantage of the notion of index is that it is invariant under
unitary time developments which include a fundamental phenomenon such
as squeezing. Another advantage of the index idea lies in suggesting a close
analogy
between the problem of quantum phase operator with a non-trivial index
as in (\ref{rel5}) and chiral anomaly in gauge theory, which is related
to the Atiyah-Singer index theorem.  This was emphasized in Ref
\cite{fuji}.  From an anomaly view point, it is not surprising to have
an anomalous identity

\begin{equation}
C( \varphi)^{2} + S( \varphi)^{2} = 1 - \frac{1}{2} |0><0|
\end{equation}

\noindent and an anomalous commutator

\begin{equation}
[C( \varphi) , S( \varphi)] =  \frac{1}{2i} |0><0|
\end{equation}

\noindent for the modified cosine and sine operators defined  in terms
of $\expon^{i \varphi}$ in (\ref{rel4}) \cite{suss}

\begin{eqnarray}
C( \varphi) &\equiv& \frac{1}{2} \{ \expon^{i \varphi} +
(\expon^{i \varphi})^{\dagger}
\}, ~ ~ ~\nonumber \\
S( \varphi) &\equiv& \frac{1}{2i} \{ \expon^{i \varphi} -
(\expon^{i \varphi})^{\dagger}
\}
\end{eqnarray}
The notion of index is also expected to be invariant
under a continuous deformation such as the quantum deformation of the
oscillator algebra as long as the norm of the Hilbert space is kept
positive definite.

\section{Q-deformation}
The purpose of the present note is to analyze in detail the behavior
of the index relation under the quantum deformation of the oscillator
algebra \cite{mac,bied}:

\begin{eqnarray}
{}~ [ a , a^{\dagger} ] &=& [N + 1] - [N] \nonumber \\
{}~ [ N, a^{\dagger} ] &=& a^{\dagger} \nonumber \\
{}~ [ N, a ] &=& - a \label{alg}
\end{eqnarray}

\noindent where

\begin{equation}
[N]  \equiv { \frac{q^{N} - q^{-N}}{q - q^{-1}}}.
\end{equation}

\noindent The parameter $q$ stands for the deformation parameter, and
one recovers the conventional algebra in the limit $q \rightarrow 1$.
The quantum deformation (\ref{alg}) is known to satisfy the Hopf
structure \cite{yan, oh}. The algebra(\ref{alg}) accomodates a Casimir operator
defined by \cite{oh}
\begin{equation}
c = a^{\dagger}a -[N]
\end{equation}
which plays an important role in the following.

For a real positive $q$, we may adopt the conventional Fock state
representation of the algebra (\ref{alg}) defined by \cite{mac, bied}:

\begin{eqnarray}
c |0> & = & 0 \nonumber\\
a |0> & = & 0 \nonumber \\
< 0 | 0 > & = & 1 \nonumber \\
N |k> & = & k|k> \label{rel20} \\
|k> & = & \frac{1}{\sqrt{[k]!}} (a^{\dagger})^{k} |0 > \nonumber \\
a|k > & =&  \sqrt{[k]}|k - 1 >, ~ ~ ~
a^{\dagger} | k >  =  \sqrt{[k + 1]} | k + 1>, ~ ~ ~  \nonumber
\end{eqnarray}

\noindent Here we have abbreviated $|k>_{q}$ by $|k>$.
For a positive $q$, one thus obtains a representation

\begin{equation}
a = |0 ><1|\sqrt{[1]} +  |1 ><2|\sqrt{[2]} + |2 ><3|\sqrt{[3]} + \cdots
\label{eqn21}
\end{equation}

\noindent which satisfies the index condition (\ref{rel2}). The phase
operator $\expon^{i \varphi}$  is defined by \cite{chiu}

\begin{eqnarray}
\expon^{i \varphi} & = & \frac{1}{\sqrt{[N+1]}}a\nonumber\\
                   & = & |0 ><1| +  |1 ><2| + |2 ><3| + \cdots   \label{rel22}
\end{eqnarray}

\noindent so that the relation $a = \expon^{i \varphi} \sqrt{N}$ holds.
Evidently, expression (\ref{rel22}) has the same form as that of Susskind and
Glogower in \cite{suss}, namely not only the index but also the
explicit form of $\expon^{i \varphi}$ itself remains invariant under
quantum deformation.

If one extends the range of the  deformation parameter $q$ to
complex numbers, which is consistent only for $|q| = 1$, one finds more
interesting
possibility. For previous discussions of this case from a finite
dimensional cyclic representation, see papers in \cite{floratos}.

For a complex $q = \exp (2\pi i\theta)$ with a real $\theta$, we adopt
the following explicit matrix representation \cite{rideau} of the algebra
(\ref{alg})

\begin{eqnarray}
a & = & \sum_{k=1}^{\infty} \sqrt{[k - n_{0}] + [n_{0}]} |k - 1><k|
\nonumber \\
a^{\dagger} & = & \sum_{k=1}^{\infty} \sqrt{[k + 1 - n_{0}] + [n_{0}]}
|k + 1><k| \label{unit}\\
N & = & \sum_{k = 0}^{\infty} (k - n_{0}) |k> <k| \nonumber  \\
c & = & [n_{0}] ~ = ~ \frac{1}{|\sin 2 \pi \theta|} \nonumber
\end{eqnarray}

\noindent Here the ket states $|k>$ stand for column vectors

\begin{equation}
|0> ~ = ~ \left( \begin{array}{c}
1 \\ 0 \\ 0 \\ \vdots \\ \vdots \\ \end{array} \right), ~ ~
|1> ~ = ~ \left( \begin{array}{c}
0 \\ 1 \\ 0 \\ \vdots \\ \vdots \\ \end{array} \right), ~ ~
|2> ~ = ~ \left( \begin{array}{c}
0 \\ 0 \\ 1 \\ \vdots \\ \vdots \\ \end{array} \right), ....
\end{equation}

\noindent and the bra states stand for row vectors. The representation
(\ref{rel20}) may also be included in this matrix representation by
letting  $n_{0} = 0$ and $c=0$.  In eq(\ref{unit}) the Casimir operator $c$ for
the
algebra (\ref{alg}) is chosen so that $a^{\dagger} a > 0$ and the
absence of negative norm is ensured.  We note that

\begin{eqnarray}
[k - n_{0}] & = & \frac{\sin 2 \pi (k - n_{0}) \theta}{\sin 2 \pi
\theta} \nonumber \\
& = & - \frac{\cos (2 \pi k \theta)}{|\sin 2 \pi \theta|} \\
& \leq & \frac{1}{|\sin 2 \pi \theta|} \nonumber
\end{eqnarray}

\noindent if one chooses $n_{0}$ as in (\ref{unit}),

\begin{equation}
[n_{0}] ~ = ~ \frac{\sin (2 \pi n_{0} \theta)}{|\sin 2 \pi \theta|} ~ =
{}~ \frac{1}{|\sin 2 \pi \theta|}
\end{equation}

\noindent The argument of the square root in (\ref{unit}) is thus
non-negative. This means that we have managed to overcome the problem
of negative norm for $q=exp(2\pi i\theta)$.
For irrational $\theta$

\begin{equation}
[k - n_{0} ] + [n_{0}] = 0
\end{equation}

\noindent only if $k = 0$.

We thus have the kernels, ker $a = \{ |0 > \}$  and ker $ a^{\dagger} =
\mbox{\rm empty} $, and the index condition

\begin{equation}
\dml a - \dml a^{\dagger} = 1
\end{equation}

\noindent for a positive $q$ or $q = \exp( 2 \pi i \theta)$ with an
irrational $\theta$: this index relation allows only the non-hermitian
phase operator defined in (\ref{rel22}), namely

\begin{eqnarray}
\expon^{i \varphi} & = & \frac{1}{\sqrt{[N+1] + [n_{0}]}} a \label{rel29}\\
                   & = & |0 ><1| +  |1 ><2| + |2 ><3| + \cdots   \nonumber
\end{eqnarray}

\noindent This expression together with $[N + 1] + [n_{0}] \neq 0$
shows that $\expon^{i \varphi}$ and $a$ carry the same index, namely a
unit index.

We next examine the representation (\ref{unit}) for a rational
$\theta$. To be specific, we consider the case $\dis q = \exp(\frac{2 \pi
i}{(s + 1)})$, i.e., $\dis \theta = \frac{1}{s + 1}$ with a positive
integer $s$ greater than one.  One then obtains

\begin{equation}
[s + 1] ~ = ~ \frac{q^{s + 1} - q^{- s - 1}}{q - q^{-1}} ~ = ~ 0
\end{equation}

\noindent In this case, the representation (\ref{unit}) becomes

\begin{eqnarray}
a & =  ~ & \sqrt{[1 - n_{0}] + [n_{0}]} |0 >< 1| + \cdots + \sqrt{[s -
n_{0}] + [n_{0}]}|s - 1>< s| \nonumber \\
& & +  \sqrt{[1 - n_{0}] + [n_{0}]} |s + 1 ><s + 2 | + \cdots + \sqrt{[s -
n_{0}] + [n_{0}]}|2s>< 2s + 1| \nonumber \\
& & +  \cdots \nonumber \\
N & = & ~ ( - n_{0}) |0 >< 0| + (1 - n_{0}) |1 >< 1| +  \cdots + (s -
n_{0})|s >< s| \label{rel31} \\
& & + (s + 1 - n_{0}) |s + 1 >< s + 1| +   \cdots + ( 2s + 1 -
n_{0})|2s + 1 >< 2s + 1| \nonumber \\
& & + \cdots \nonumber \\
c & = &  [n_{0}] = \dis{\frac{\sin(\frac{2 \pi n_{0}}{s + 1})}{\sin (\frac{2
\pi }{s + 1})}} = {\dis \frac{1}{\sin (\frac{2 \pi}{s + 1})}} \nonumber
\end{eqnarray}

\noindent where $a^{\dagger}$ is given by the hermitian conjugate of $a$
and one may choose $n_{0} = {\dis \frac{s + 1}{4}}$.

One may look at the representation (\ref{rel31}) from two different
view points.  One way is to regard it reducible into an infinite set of
irreducible $(s + 1)$- dimensional representation specified by the
eigenvalue of the Casimir operator $ c = [n_{l}] ~  (= [n_{0}]) ~ $ where

\begin{eqnarray}
n_{l} & = & n_{0} - l (s + 1) \nonumber \\
& = & \frac{1}{4}(s + 1) - l (s + 1)
\end{eqnarray}

\noindent with $l = 0, 1, 2, \cdots $.  We note that $- n_{l}$ stands
for the lowest eigenvalue of $N$.  In this case, the basic Weyl block
is given by

\begin{eqnarray}
a_s & = & \sqrt{[1 - n_{0}] + [n_{0}]} |0 >< 1| + \cdots + \sqrt{[s -
n_{0}] + [n_{0}]}|s - 1>< s| \nonumber \\
a_s^{\dagger} &= & (a_s)^{\dagger} \nonumber \\
N_s & = & ( - n_{0}) |0 >< 0| + (1 - n_{0}) |1 >< 1| +  \cdots + (s -
n_{0})|s >< s| \nonumber \\
c & = &  [n_{0}] = {\dis \frac{1}{\sin (\frac{2 \pi}{s + 1})}} \label{rel33}
\end{eqnarray}

\noindent and other sectors are obtained by using the Casimir operator
$c = [n_l] (= [n_0])$  with the lowest eigenvalue of $N$ at $- n_{l}$, $l
= 1, 2, \cdots $.  This is the standard  representation
commonly adopted for the case
$\dis \theta = \frac{1}{(s + 1)}$.  This finite dimensional representation
inevitably leads to the index condition[5]

\begin{equation}
\dml a_s - \dml a_s^{\dagger} = 0
\end{equation}

\noindent and one may introduce the phase operator of Pegg and Barnett
in (\ref{rel7}), which is unitary $\expon^{i \phi}( \expon^{i
\phi})^{\dagger} =(\expon^{i \phi})^\dagger  \expon^{i
\phi} = 1 $ in $(s + 1)$-dimensional space.  The large $s$-limit of this
construction leads to the problematic aspects arising from index
mismatch analysed in Ref\cite{fuji}.  Also, the large $s$-limit of
(\ref{rel33}) does not lead to the standard representation (\ref{rel20})
with well-defined Casimir operator, since $\dis n_{0} = \frac{s + 1}{4}$ in
(\ref{rel33}).

Another view of the representation (\ref{rel31}), which is
interesting from an index consideration, is to regard (\ref{rel31}) as an
infinite
dimensional representation specified by the Casimir operator $c =
[n_0]$ with $- n_0$ the lowest eigenvalue of $N$.  We then have the
kernels
\begin{eqnarray}
ker ~ a & = & \{ |0>, |s + 1> |2s + 2>, \cdots \} \nonumber \\
ker ~ a^\dagger & = & \{ |s>, |2s + 1>, \cdots \}
\end{eqnarray}

\noindent and

\begin{equation}
\dml a = \infty, ~ ~ \dml a^\dagger = \infty \label{rel36}
\end{equation}

\noindent Consequently, (\ref{rel31}) corresponds to a $singular$
 point
of index theory where the notion of index becomes ill-defined:
we have no constraint on the phase operator arising from an index
consideration. In fact, one may accomodate either the non-unitary
$\expon^{i \varphi}$ in (4), which is normally associated with
$$\dml a  -  \dml a^\dagger = 1,$$
\noindent or a unitary $\expon^{i \Phi}$ defined
by

\begin{eqnarray}
\expon^{i \Phi} & = & ~ |0 ><1| +  |1 ><2| + |2 ><3| + \cdots
+ \expon^{i \phi_0} |s><0| \nonumber \\
& & + |s + 1><s + 2| + \cdots + \expon^{i \phi_1} |2s + 1><s + 1|
\nonumber \\
& & + \cdots
\end{eqnarray}

\noindent with $\phi_0$, $\phi_1$, $\cdots$, real constants; unitary
$\expon^{i \Phi}$ is normally associated with $$\dml a  -  \dml
a{^\dagger}= 0.$$
\noindent   Both of these phase operators give rise to the same
representation for $a$ in (\ref{rel31}),

\begin{eqnarray}
a & = & \expon^{i \varphi} \sqrt{[N] + [n_0]} \nonumber \\
& = & \expon^{i \Phi} \sqrt{[N] + [n_0]}
\end{eqnarray}

\noindent However, we have no more the expression in (\ref{rel29})
since $[N+1] + [n_{0}]$ can vanish.
The operator $\expon^{i \Phi}$ gives rise  to the same physical
implications as $\expon^{i \phi} $ in (\ref{rel7}) for the physical
states defined in (\ref{rel12}).

\section{Discussion and Conclusion}
We would like to summarize the implications of the above analysis.
First of all, the notion of index is well-defined for a real positive
$q$ (which includes $q =1$), and the index is invariant under a
continuous deformation specified by $q$.  The notion of index presents
a stringent constraint on the possible form of the phase operator.

For $q = \exp(2 \pi i \theta)$, the notion of index becomes subtle.
Since the rational values of $\theta$ are densely distributed among the
real values of $\theta$, one cannot define a notion of continuous
deformation for the index (i.e., $\dml a  -  \dml a^\dagger$); one
encounters
singular points associated with a rational $\theta$ almost everywhere.
Only when one regards the singular situation such as in (\ref{rel36})
as corresponding to the index relation

\begin{equation}
\dml a  -  \dml a^\dagger = 1 \label{rel39}
\end{equation}

\noindent one maintains the notion of continuous deformation.  Even in
this case, there is certain complication for $\theta \rightarrow 0$ to
reproduce the normal case of $q = 1$ if one sticks to
representation (\ref{unit}); the Casimir operator cannot
be well-defined in the limit $\theta \rightarrow 0$ as it should be.

If one formally defines the representation
\renewcommand{\thefootnote}{\fnsymbol{footnote}}{\footnote[4]{Note that
representation (\ref{rel40}) generally contains negative norm states for $q =
\exp(2 \pi i \theta)$}}

\begin{eqnarray}
a & = & \sum_{k = 1}^{\infty} \sqrt{[k]} |k - 1><k| \nonumber \\
a^\dagger & = & \sum_{k = 1}^{\infty} \sqrt{[k + 1]} |k + 1><k |
\label{rel40} \\
N & = & \sum_{k = 0}^{\infty} k |k >< k| \nonumber\\
c & = & 0 \nonumber
\end{eqnarray}

\noindent for all allowed values of $q$ and if one formally takes the
index (\ref{rel39}) even for a rational $\theta$, one can maintain the
notion of continuous deformation of the algebra and its representation.
Only in this case,  the index as well as the phase operator remain
invariant under $q$-deformation.  The standard finite dimensional
representation for $q = \exp(2 \pi i \theta)$ with a rational
$\theta$ may be interpreted that the well-defined notion of index,
 which
is supposed to be invariant under deformation, is lost for a rational
$\theta$ and the representation makes a discontinuous transition to
finite dimensional irreducible representations.

In conclusion, the notion of index, when it is well-defined, is useful
as an invariant characterization of $q$-deformation of an algebra.
In addition, we have also shown how to overcome the problem of negative norm
for $q = \exp(2 \pi i \theta)$.

\end{document}